\documentclass[aps,prb,twocolumn,groupedaddress]{revtex4}
\usepackage[pdftex]{graphicx}
\usepackage{epstopdf}

\begin{document}

\title{Broadband reconfigurable logic gates in phonon waveguides}

\author{D. Hatanaka}

\email{hatanaka.daiki@lab.ntt.co.jp}

\author{T. Darras}

\author{I. Mahboob}

\author{K. Onomitsu}

\author{H. Yamaguchi}

\affiliation{NTT Basic Research Laboratories, NTT Corporation, Atsugi-shi, Kanagawa 243-0198, Japan}

\maketitle
\hspace*{0.5em}\textbf{The high-quality-factor mechanical resonator in electromechanical systems has facilitated dynamic control of phonons via parametric nonlinear processes and paved the development of mechanical logic-elements. However the resonating element with a narrow bandwidth limits the resultant operation speeds as well as constraining the availability of nonlinear phenomena to a narrow spectral range. To overcome these drawbacks we have developed phonon waveguides in which the mechanical analogue of four-wave-mixing is demonstrated that enables the frequency of phonon waves to be converted over 1 MHz. We harness this platform to execute multiple binary mechanical logic gates in parallel, via frequency division multiplexing in a two-octave-wide phonon transmission band, where each gate can be independently reconfigured. The fidelity of the binary gates is verified via temporal measurements yielding eye diagrams which confirm the availability of high speed logic operations. The phonon waveguide architecture thus offers the broadband functionality that is essential to realising mechanical signal processors.}\\
\hspace*{1.5em}Mechanical computing was pioneered in Babbage's difference engines in 1822\cite{babbage, swade} but this platform was rendered obsolete with the emergence of the transistor in 1947\cite{alaina}. Nowadays transistor technology faces the well-known limitations of Moore's Law but interestingly, mechanical systems have resurfaced again as an alternative technology for computational hardware\cite{itrs2015}. Advanced fabrication techniques have enabled the creation of nano/micro-electromechanical systems (N/MEMS) where the mechanical element inhabits a tiny footprint, in contrast to its archaic ancestors,  and the integrated electrical circuit provides the means to manipulate mechanical vibrations i.e. acoustic phonons with miniscule power consumption\cite{roukes2001, roukes2004, ekinci_nems}. Indeed this latter prospect in particular has stimulated a flurry of activity to develop mechanical components that could be exploited in a nanomechanical computer\cite{guerra_switch, kraus_nems, badzey_memory, badzey_stochastic, masmanidis_logic, imran_memory, venstra_bistable, lee_inverter, imran_parallel, wenzler_fredkin, hatanaka_pnc, imran_register, hatanaka_ram, sklan_phonon_comp, hafiz_logic, li_meta}.\\
\hspace*{1.5em}A binary logic gate is at the heart of a digital computer which is constructed by wiring multiple transistors into a circuit. As a first step, the primary logic gates (AND, OR, NOT/XOR\cite{masmanidis_logic, lee_inverter, imran_parallel, li_meta, hafiz_logic}) were replicated using N/MEMS where binary information was encoded into either the amplitude or phase of the mechanical vibration. Subsequently logic circuits were also developed either by combing mechanical elements\cite{wenzler_fredkin} or by spectrally multiplexing a single mechanical resonance\cite{imran_parallel}. In particular this latter approach was predicated on parametric oscillations which could be activated by nonlinearly varying a parameter of the mechanical element via electrical means\cite{westra_nonlinear, karabalin_nonlinear, karabalin_nonlinear2, eichler_nonlinear, venstra_stochastic,yamaguchi_nonlinear}. Indeed the dynamic control of phonons in this regime provides a universal platform in which logic-circuits\cite{imran_parallel,li_meta}, memory\cite{imran_memory}, switches\cite{hatanaka_pnc,li_meta} and a shift register\cite{imran_register} can be implemented namely all the key components one would need to build a nanomechanical computer. In spite of the success of this approach it comes at the cost of a narrow operation bandwidth due to the high quality factor of the mechanical resonance which while enhancing the fidelity of the nonlinear modulation at the heart of the parametric oscillations, it limits the processing speed of the resultant logic operations. This in turn suppresses the processing power of such a mechanical computer to the point where it becomes an unviable alternative to transistor based computation.\\
\hspace*{1.5em}To overcome this drawback we demonstrate the above concepts using a phonon waveguide (WG) which is composed from an array of vibrating membranes that are strongly coupled and it results in their individual resonances merging to give rise to a broadband transmission band\cite{hatanaka_pnc, hatanaka_njp}. The broad transmission band can still host a Duffing nonlinearity, corresponding to a cubic term of the mechanical vibration amplitude, which can be activated due to the strongly confined vibrations within the membrane array and by exciting the WG from the highly efficient piezoelectric transducers that are integrated directly into the membranes. This nonlinearity is harnessed to implement the mechanical analogue of four-wave-mixing (FWM)\cite{agrawal,foster_fwm,ferrera_fwm} which enables AND, OR and XOR logic gates to be executed with the speeds exceeding previous mechanical resonator-based demonstrations.

\section{Results}
\hspace*{-1em}\textbf{Phonon waveguides.}\hspace*{1em}The phonon WG, shown in Fig. 1(a), consists of a one-dimensional array of coupled membrane mechanical resonators made from a GaAs/AlGaAs heterostructure\cite{hatanaka_pnc, hatanaka_njp}. Periodically arrayed air-holes are defined on the 1mm long WG which are used to suspend the membranes as described in Methods. The experiments were performed using two WGs having (${\it w}$, ${\it a}$) = (34, 8) and (34, 12) in ${\mu}$m, where ${\it w}$ and ${\it a}$ are waveguide width and hole pitch respectively, and hereafter they are labelled WG 1 and WG 2 respectively. The piezoelectric effect in the GaAs-based structure enables phonon vibrations to be excited by applying an alternating voltage to one of the electrodes located on both edges of the WG. The resultant phonons travel down the WG and are measured at the other edge via optical interferometry. All the experiments were performed at room temperature and in a high vacuum (2 ${\times}$ 10$^{-4}$ Pa).\\
\hspace*{1em}First the spectral response of the phonon transmission in WG 1 is investigated by exciting and measuring phonon waves at the left and right edges respectively. The phonon vibrations are observed from 1.5 to 8.2 MHz as shown in Fig. 1(b) where equidistant peaks from Fabry-Perot (FP) resonances can also be seen. Although the existence of the periodic air-holes can give rise to a phonon bandgap around 6.5 MHz in this device, the existence of another phonon branch at this frequency obscures it in the spectral domain thus permitting a continuous band of mechanical vibrations over two-octaves wide to be accessed in the phonon WG\cite{hatanaka_njp}.

\vspace{0.2cm}

\hspace*{-1em}\textbf{Four wave mixing.}\hspace*{1em}The WG structure can effectively confine phonon waves within the suspended membranes and in combination with the highly potent piezotransducers. It enables nonlinearities to be induced into the elastic restoring force\cite{eichler_nonlinear,yamaguchi_nonlinear}. The strong excitation of a pump phonon wave with frequency ${\it f}_{\rm p}$ activates a cubic, namely Duffing, nonlinearity in the elastic restoring force which enables FWM to be executed. This allows an incident signal phonon wave with frequency ${\it f}_{\rm s}$ to interact with the pump wave which in turn generates a new idler phonon wave with frequency ${\it f}_{\rm i}=2{\it f}_{\rm p}-{\it f}_{\rm s}$. In contrast to the parametric frequency modulation of the restoring force previously employed to implement three-wave-mixing (TWM) in a resonating electromechanical system to execute logic gates and circuits\cite{imran_parallel}, the FWM process is suited to the phonon WG as the frequencies of the phonon waves involved in this process are more closely spaced which enables phase matching between the different waves to be readily satisfied in addition to energy conservation. \\
\hspace*{1em}In order to confirm the availability of FWM in the phonon WG architecture, the spectral response of WG 2 is monitored by simultaneously activating a weak signal (blue) and a strong pump wave (red). The results of this measurement shown in the lower panel of Fig. 2 reveal the generation of an idler wave (green) from the FWM interaction between the signal and pump waves. Moreover the frequency of the idler wave can be continuously tuned over 1 MHz as the signal excitation is adjusted indicating the availability of broadband frequency conversion in contrast to the narrow band frequency conversion via TWM previously demonstrated in a resonating mechanical system (see Supplementary Note 1 and Supplementary Fig. 1). Note that WG 2 hosts equally spaced FP resonances in the phonon transmission band, as shown in the upper panel of Figs. 2, and therefore, the idler amplitude is enhanced when it overlaps with them corresponding to perfect phase matching in the FWM process\cite{ferrera_fwm,morichetti_crow,kippenberg_comb} as detailed in Supplementary Note 2 and Supplementary Fig. 2.\\

\vspace{0.2cm}

\hspace*{-1em}\textbf{Logic operations in the phonon WG.}\hspace*{1em}The mechanical FWM can be harnessed to implement the AND, OR and XOR fundamental logic gates. In this approach two signal waves at ${\it f}_{\rm s1}$ and ${\it f}_{\rm s2}={\it f}_{\rm s1}+\Delta$ are used to encode binary logical inputs and the generated idler wave yields the binary output with the presence (absence) of a phonon vibration being defined as 1 (0). By adjusting the number, frequency and phase of the pump waves used to activate the mechanical FWM, a range of idler waves can be generated whose signal wave dependence enables AND, OR and XOR gates to be created in the phonon WG. In practice the primary logic gates are implemented in one of four spectral regions defined in WG 1, henceforth labelled G1, G2, G3, G4 and detailed in Fig. 1(b), yielding a Boolean logic gate array in frequency space.\\
\hspace*{1em}In order to implement an AND gate, a pump wave (${\it f}_{\rm p}$) and two signal waves (${\it f}_{\rm s1}$ and ${\it f}_{\rm s2}$) are simultaneously injected into the WG resulting in two idler waves being generated at ${\it f}_{\rm i1}^{(1)}=2{\it f}_{\rm p}-{\it f}_{\rm s1}$ and ${\it f}_{\rm i2}^{(1)}=2{\it f}_{\rm p}-{\it f}_{\rm s2}$ from FWM. These idler waves then serve as seeds for further FWM yielding secondary idler waves at ${\it f}_{\rm i12}^{(2)}=2{\it f}_{\rm i1}^{(1)}-{\it f}_{\rm i2}^{(1)}$ and ${\it f}_{\rm i21}^{(2)}=2{\it f}_{\rm i2}^{(1)}-{\it f}_{\rm i1}^{(1)}$. The secondary idlers can only be observed when both signal waves are injected into the WG thus naturally leading to an AND gate (see Supplementary Note 3.1). In the middle panel of Fig. 3(a), this gate is experimentally implemented in region G1 with signal waves ${\it f}_{\rm s1}$ = 2.830 MHz and ${\it f}_{\rm s2}$ = 2.880 MHz namely ${\Delta}$ = 50 kHz (blue), pump wave ${\it f}_{\rm p}$ = 3.254 MHz (red) and the secondary idler waves (green) are observed at ${\it f}_{\rm i12}^{(2)}$ = 3.728 MHz and ${\it f}_{\rm i21}^{(2)}$ = 3.578 MHz only when both signals are activated. On the other hand, when either of the two signals is not excited, no secondary idler waves are generated. Thus the secondary idler response corresponds to an AND gate which can be executed between 3.35-3.75 MHz as shown in the right panel of Fig. 3(a) (see Supplementary Fig. 3 for more details). \\
\hspace*{1em}Next to implement an OR gate, spectrally degenerate idler waves are generated from two pump waves at ${\it f}_{\rm p1}$ and ${\it f}_{\rm p2}={\it f}_{\rm p1}+\Delta/2$ in addition to the two signal waves (${\it f}_{\rm s1}$ and ${\it f}_{\rm s2}$). This results in four idler waves of which two can be made degenerate at frequency ${\it f}_{\rm i11}=2{\it f}_{\rm p1}-{\it f}_{\rm s1}=2{\it f}_{\rm p2}-{\it f}_{\rm s2}={\it f}_{\rm i22}$ by appropriately tuning ${\it f}_{\rm p2}$ (see Supplementary Note 3.2). Consequently this enables the degenerate idlers to be observed when either or both signals are injected into the WG yielding an OR gate. Experimentally two signals ${\it f}_{\rm s1}$ = 3.825 MHz and ${\it f}_{\rm s2}$ = 3.881 MHz (blue) and two pumps ${\it f}_{\rm p1}$ = 4.347 MHz and ${\it f}_{\rm p2}$ = 4.375 MHz namely ${\Delta}$ = 56 kHz (red) are injected into G2 which results in output idler waves at ${\it f}_{\rm i11}={\it f}_{\rm i22}$ = 4.869 MHz (green) as shown in the middle panel of Fig. 3(b). This measurement reveals the OR gate can be successfully executed between 4.5-4.9 MHz as shown in the right panel of Fig. 3(b), when the signal frequencies are swept, from the degenerate idlers generated by the precisely tuned dual pump excitations (see Supplementary Fig. 4 for more details).\\
\hspace*{1em}Finally to realise the XOR gate, destructive interference between the degenerate idler waves, used for the OR gate, is utilised. Specifically two pump waves tuned to yield degenerate idlers, but now with a ${\pi}$/2 phase difference, are applied to the WG which causes the idler waves to interfere and cancel out resulting in the XOR gate operation (see Supplementary Note 3.3 and Supplementary Fig. 5). In the experiment, two pumps ${\it f}_{\rm p1}$ = 5.356 MHz and ${\it f}_{\rm p2}$ = 5.390 MHz (i.e. ${\Delta}$ = 67.5 kHz) with a relative ${\pi}$/2 phase shift (red and purple in the middle panel of Fig. 3(c)) and two signals ${\it f}_{\rm s1}$ = 4.860 MHz and ${\it f}_{\rm s2}$ = 4.928 MHz are injected into region G3 of WG 1. This results in the degenerate idler wave at ${\it f}_{\rm i11}={\it f}_{\rm i22}$ = 5.852 MHz  being eliminated, as shown in the middle panel of Fig. 3(c), whereas excitation of either one of the two signals yields an idler wave output. The XOR output can also be observed between 5.45-5.9 MHz in the phonon WG when the signal wave frequencies are swept as shown in the right panel of Fig. 3(c) (see Supplementary Fig. 6 for more details).\\
\hspace*{1em}The AND, OR and XOR gates can be executed in all 4 spectral regions G1-G4 in the phonon WG by simply adjusting the pump conditions. More importantly, the idler outputs can be used as signal inputs for further logic operations that would enable the realisation of a series of logic gates and construction of complex logic circuits. Consequently multiple and reconfigurable logic gates via FWM can be implemented over the broad transmission band in the phonon WG.\\

\vspace{0.2cm}

\hspace*{-1em}\textbf{High speed mechanical logic gates.}\hspace*{1em}The broad transmission band in the phonon WG also offers the prospect of high speed logic gates. To that end the AND gate was realised in region G3 in WG 1 with two signal waves at ${\it f}_{\rm s1}$ = 5.010 MHz and ${\it f}_{\rm s2}$ = 5.077 MHz which were rapidly amplitude modulated by a pseudo-random bit sequence (PRBS) at 3 kb/s and a continuous wave (CW) pump at ${\it f}_{\rm p}$ =5.322 MHz that were simultaneously injected into the WG. The output idler at ${\it f}_{\rm i12}^{(2)}$ = 5.702 MHz is only generated during the presence of both signal inputs, as shown in the right panel of Fig. 4(a), and thus it can successfully encode the AND gate at 3 kb/s. Additionally the OR gate was realised in region G1 in WG 1 with two signal waves injected at ${\it f}_{\rm s1}$ = 3.010 MHz and ${\it f}_{\rm s2}$ = 3.060 MHz (i.e. ${\Delta}$ = 50 kHz) which again were rapidly amplitude modulated by the PRBS at 3 kb/s and two CW pumps injected at ${\it f}_{\rm p1}$ = 3.279 MHz and ${\it f}_{\rm p2}$ = 3.304 MHz. Now the output idler at ${\it f}_{\rm i11}={\it f}_{\rm i22}$ = 3.548 MHz is only eliminated if both signal inputs are absent as shown in the right panel of Fig. 4(b) and thus it can also encode the OR gate at high speed. It should be noted that the OR idler's amplitude varies when only ${\it f}_{\rm s2}$ is active in contrast to when ${\it f}_{\rm s1}$ or both ${\it f}_{\rm s1}$ and ${\it f}_{\rm s2}$ are activated and is due to idler conversion efficiencies being different for the two pumps.\\
\hspace*{1em}The fidelity of the high speed AND and OR logic gates can be evaluated via the eye pattern acquired from the above measurement protocol with multiple runs of the PRBS at 1 kb/s and is shown in Figs. 4(c) in the left and right panels respectively. The eye openings indicate that the AND gates can be successfully executed even with the presence of multiple input phonon waves that yield a rich idler spectrum. Note that the eye opening is less clearly defined for the OR gate as consequence of the differing idler conversion efficiencies between the two pumps as detailed above.\\

\vspace{0.2cm}

\hspace*{-1em}\textbf{Parallel logic gates.}\hspace*{1em}Another consequence of the broad transmission band in the WG is the possibility to execute logic gates in parallel in the different gate regions G1-G4 as detailed in Fig. 1(b) by frequency-division multiplexing the input signal and pump waves. This concept is demonstrated in WG 1 with the AND and OR gates being implemented simultaneously. Specifically two signals waves ${\it f}_{\rm s1}$ = 2.940 MHz and ${\it f}_{\rm s2}$ = 2.990 MHz, which are randomly amplitude modulated, and one CW pump ${\it f}_{\rm p1}$ = 3.254 MHz are injected in the WG to generate the output from a secondary idler at ${\it f}_{\rm i12}^{\rm (2)}$ = 3.618 MHz for the AND gate in G1 (see the left insets of Fig. 5). Simultaneously two additional signal waves ${\it f}_{\rm s3}$ = 6.020 MHz and ${\it f}_{\rm s4}$ = 6.110 MHz with the same random amplitude modulation and two CW pumps ${\it f}_{\rm p2}$ = 6.315 MHz and ${\it f}_{\rm p3}$ = 6.360 MHz are injected into the WG to generate the idler at ${\it f}_{\rm i23}={\it f}_{\rm i34}$ = 6.610 MHz for the OR gate in G4 (see the left insets of Fig. 5). The temporal response of the amplitudes of the output idlers ${\it f}_{\rm i12}^{\rm (2)}$ and ${\it f}_{\rm i23}$ (or ${\it f}_{\rm i34}$), acquired concurrently and shown in the right panels of Fig. 5, confirms that they can successfully encode AND and OR gates simultaneously. Consequently this result indicates that frequency-division-multiplexing can be applied to the signal and pump waves injected into the phonon WG to realise a spectral logic array in which Boolean logic gates can be executed in parallel.\\

\section{Discussion}
\hspace*{0em}The key merit of the phonon WG is its broad transmission band which allows the operation speed of mechanical logic gates to be increased up to ${\sim}$ 3 kb/s that is significantly faster than conventional resonance based mechanical systems. In spite of this advance, and from the view point of practical applications, this speedup is still insufficient and several orders of magnitude of further enhancement is needed. \\
\hspace*{1em}The current device is confined to the 1.5-8.2 MHz operation frequency of the transmission band and the switching speed is limited by the 8 kHz bandwidth of the FP resonances. Both of these properties need to be improved before the phonon WG architecture can be considered as a practical system for information signal processing. First to increase the frequency of the transmission band requires further miniaturisation of the WG down to submicron scales for hypersonic phonons which in principle will concomitantly be accompanied by a wider transmission band. Second, a flat transmission band without FP resonances will enable the switching speed to be increased up to the bandwidth of the transmission band. The low conversion efficiencies likely to emerge due to the absence of FP resonance will then need to be compensated for by careful engineering of the phonon dispersion relations that would permit the phase matching needed for FWM.\\
\hspace*{1em}A potential platform for hypersonic phonons could be accessed from two-dimensional phononic crystals (2D PnCs) where a line defect in a 2D periodic structure can form a phonon waveguide in which GHz phonons can be spatially confined and guided in a multi-GHz wide transmission band\cite{maldovan_nature, benchabane_pnc1, mohammadi_pnc1, mohammadi_pnc2, otsuka_pnc, benchabane_pnc2, balram_optomecha}. The prospect of electrically controlling hypersonic phonons in 2D PnCs offers the possibility of frequency-division-multiplexing a large number of logic gates in parallel all of which could be operated at Gb/s by applying the techniques demonstrated in this study. Furthermore the hypersonic PnC WG would also not suffer from performance limitations due to air damping as hypersound vibration amplitudes approach the mean free path of intermolecular distance in air thus enabling this architecture to be implemented in normal environmental conditions\cite{li_nems}.\\
\hspace*{1em}A phonon WG is developed in which frequency-division-multiplexed logic gates via mechanical FWM are demonstrated which are not only reconfigurable on the fly they can also be operated at kHz speeds. The prospect of exploiting these techniques in hypersonic structures paves the way towards novel phonon logic devices which in their ultimate evolution could be operated with low power consumption, thus offering the ideal realisation of environmentally friendly information technology.\\

\vspace{0.5cm}
\noindent{\textbf{{\normalsize METHODS}}}\\
{\small The phonon WGs were constructed from a GaAs/AlGaAs heterostructure by isotopically and selectively etching a 3 ${\mu}$m thick Al$_{\rm 0.65}$Ga$_{\rm 0.35}$As sacrificial layer through the periodic air-holes that were defined through a 5 nm GaAs layer on top of 95 nm Al$_{\rm 0.27}$Ga$_{\rm 0.73}$As and 100 nm n-GaAs layers with HF(5$\%$):H$_{\rm 2}$O(95$\%$) solution. The 40 min selective etching time also determined the WG's width to be 34 ${\mu}$m. The 80 nm thick Au gates were located on both edges of the WGs and they formed the top electrode of the piezoelectric stack that was used to activate the phonon waves in the WG.\\
The application of alternating electric voltage from a signal generator (NF Wavefactory 1974) to one of the electrodes induces bending mechanical vibrations in the WG due to the piezoelectric effect. The resultant vibration travels down the WG and is optically detected in a He-Ne laser Doppler interferometer (Neoark MLD-230V-200). The spectral responses in Figs. 1(b), 2 and 3(a)-(c) are obtained by demodulating the output from the interferometer in a vector signal analyzer (HP 89410A) or a lock-in amplifier (Zurich Instruments UHFLI). The temporal responses in Figs. 4 and 5 are obtained by amplitude modulating the signal generator with 2$^{\rm 14}$-1 PRBS non-return-to-zero from a pulse pattern generator (Keysight Technologies 81110A) and demodulating the interferometer in a lock-in amplifier followed by an oscilloscope (Keysight Technologies DSO9104H).}\\

\vspace{0.2cm}
\hspace*{-0.35cm}\textbf{{\small Supplementary Information}} {\small is linked to the online version of the paper at www.nature.com/nature.}\\
\\
\textbf{{\small Acknowledgements}} {\small The authors are grateful to Y. Ishikawa for growing the heterostructure and K. Nozaki for helping with the eye pattern measurements. This work is partly supported by MEXT Grant-in-Aid for Scientific Research on Innovative Areas “Science of hybrid quantum systems" (Grant No. JP15H05869).}\\
\\
\textbf{{\small Author Contributions}} {\small D.H. fabricated the  sample and D.H. and T.D. performed measurements and data analysis. K.O. co-fabricated the GaAs/AlGaAs heterostructure. I.M. and D.H. wrote the paper and H.Y. planned the project.}\\
\\
\textbf{{\small Author Information}} {\small Reprints and permissions information is available at www.nature.com/reprints. The authors declare no competing financial  interests. Readers are welcome to comment on the online version of this article at www.nature.com/nature. Correspondence and requests for materials should be addressed to D.H. (hatanaka.daiki@lab.ntt.co.jp).}


\begin{figure*}[t]
\begin{center}
\vspace{-2.5cm}\hspace{0cm}
\includegraphics[scale=0.55]{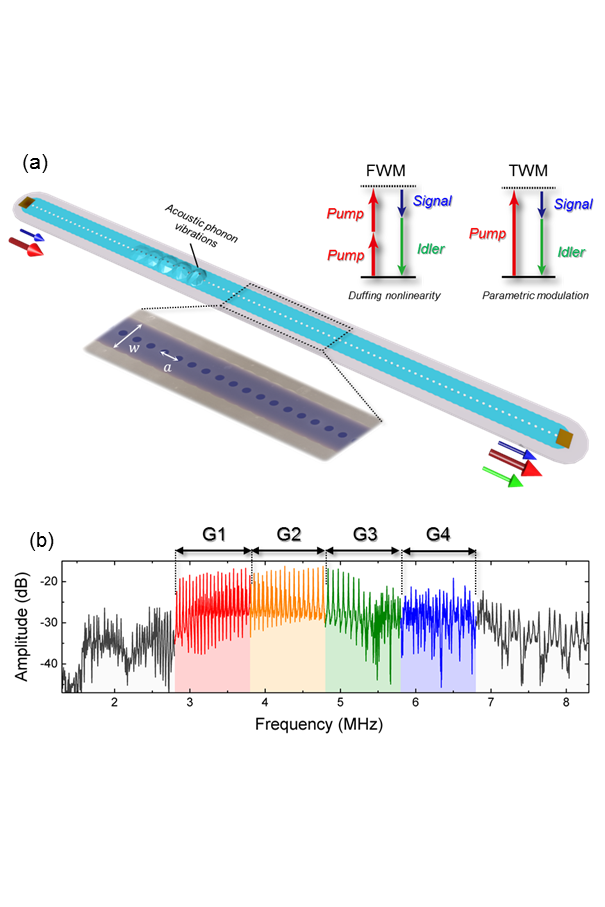}
\vspace{-2cm}
\caption{\textbf{An electromechanical phonon waveguide.} (${\bf a}$) A schematic of the FWM process in the electromechanical phonon WG.  The individual membranes (light blue) in the WG are suspended above the substrate and isolated from their surroundings (grey) thus enabling acoustic phonon vibrations to be confined and guided. The left inset shows a false-colored electron microscope image of the device indicating the WG width (${\it w}$) and the hole pitch (${\it a}$). The right insets show energy diagrams for the FWM, investigated in this report, and TWM previously employed to execute mechanical logic circuits. In the FWM process, a strong pump wave (red arrow) activates a third-order term in the restoring force in the phonon WG and the simultaneous excitation of a signal wave (blue arrow) stimulates the emission of a new idler wave (green arrow). (${\bf b}$) The phonon transmission spectrum in WG 1 when excited from the left edge with 1 V$_{\rm rms}$ and measured at the right edge via optical interferometry in a vector signal analyser. The phonon band is divided into four 1-MHz spectral regions, labelled G1, G2, G3 and G4 and highlighted in red, yellow, green and blue respectively, in order to demonstrate frequency-division-multiplexed logic gates detailed below.}
\label{fig 1}
\vspace{-0.5cm}
\end{center}
\end{figure*}

\begin{figure*}[t]
\begin{center}
\vspace{0cm}\hspace{0cm}
\includegraphics[scale=0.6]{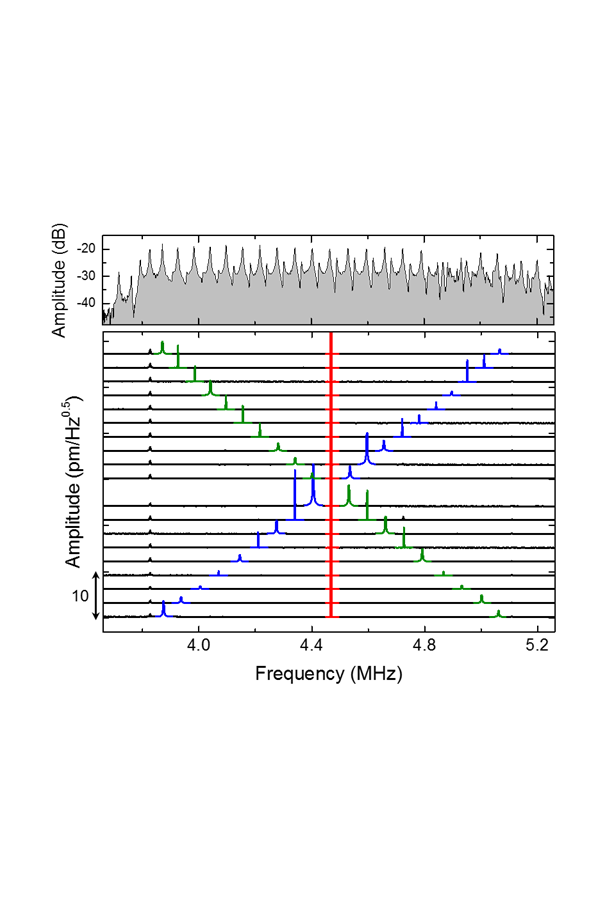}
\vspace{-3.0cm}
\caption{\textbf{Frequency conversion of phonon waves via FWM.} The spectral response of phonon WG 2 when excited with a fixed pump excitation at 4.468 MHz and an amplitude of 1 V$_{\rm rms}$ (red) whilst simultaneously the signal excitation is swept from 3.8 to 5.1 MHz with an amplitude of 0.35 V$_{\rm rms}$ (blue). Note that the idler (green) generated from FWM shifts from high to low frequencies as the signal is swept from low to high frequencies in order to conserve energy. The upper panel shows the frequency response of WG 2 when excited from the left edge with a 1 V$_{\rm rms}$ amplitude and measured at the right edge.}
\label{fig 2}
\vspace{-0.5cm}
\end{center}
\end{figure*}

\begin{figure*}[t]
\begin{center}
\vspace{-0cm}\hspace{6cm}
\includegraphics[scale=0.85]{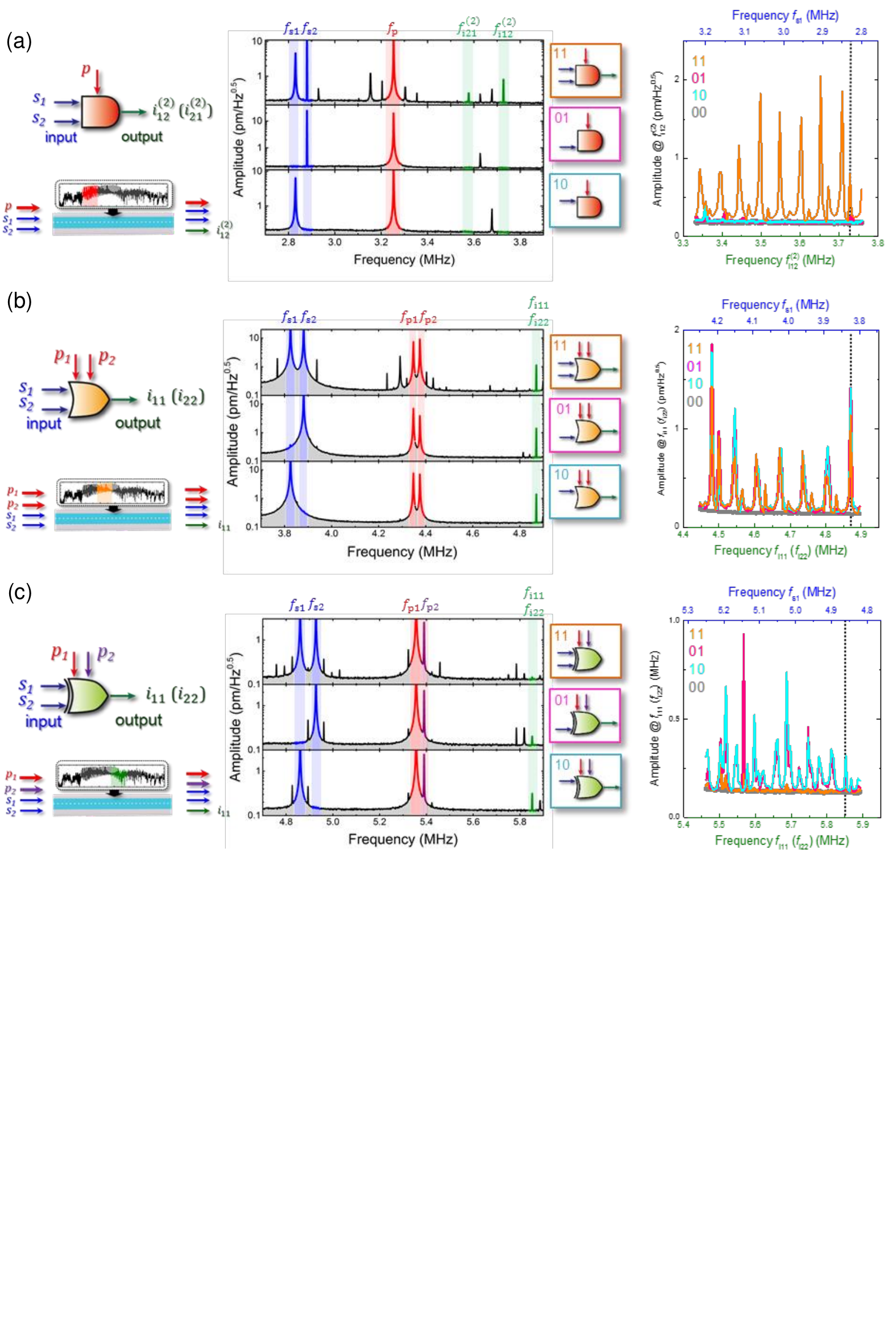}
\vspace{-6.5cm}
\caption{\textbf{Frequency-division-multiplexed logic gates in the phonon WG.} The output spectra of (${\bf a}$) AND,  (${\bf b}$) OR and (${\bf c}$) XOR logic gates executed in spectral windows G1, G2 and G3 in WG 1 respectively as detailed in Fig. 1(b). Left: The input/output configurations of the logic gate being executed via the phonon excitations in the WG where the signal, pump and idler are coloured blue, red (or purple to imply a phase shift) and green respectively throughout. Middle: All the logic gates are executed as a function of signal input configurations in the WG where signal 1 (${\it f}_{\rm s1}$) and signal 2 (${\it f}_{\rm s2}$) being on and off respectively yields the binary input 10, ${\it f}_{\rm s1}$ = off ${\it f}_{\rm s2}$ = on yields 01, both signals being on gives 11 and the corresponding input/output configurations of the logic gates are shown in the right insets. Note that numerous other peaks are generated in the WG when all the input excitations are activated which corresponds to different mixing combinations between the signal, pump and idler waves. Right: The output logical idlers can be generated over a broad range of frequencies as the signal frequencies are swept for the different input configurations of 00 (grey), 10 (light blue), 01 (pink) and 11 (orange). The dotted line indicates the idlers described in the middle panel.}
\label{fig 3}
\vspace{-0.5cm}
\end{center}
\end{figure*}

\begin{figure*}[t]
\begin{center}
\vspace{-0cm}\hspace{0cm}
\includegraphics[scale=0.8]{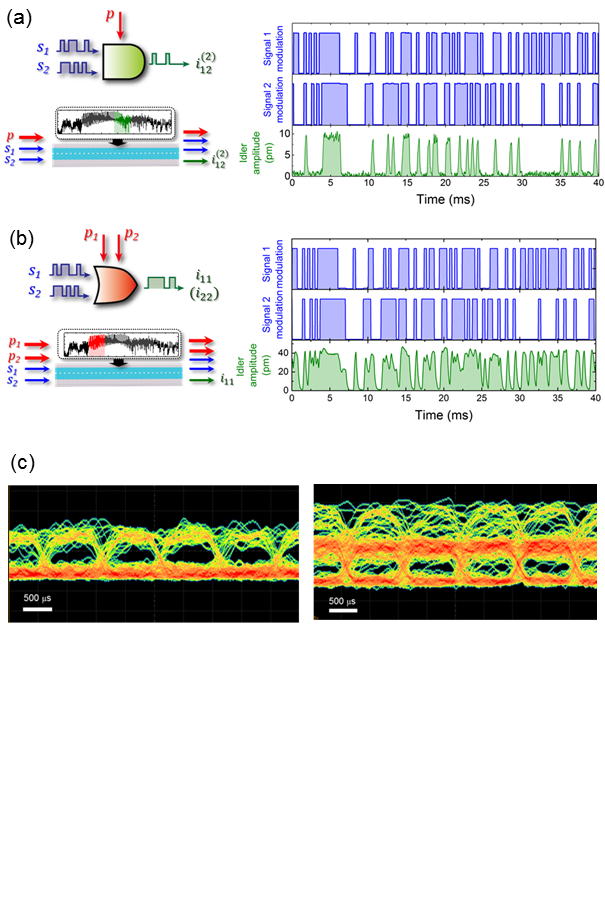}
\vspace{-5.5cm}
\caption{\textbf{High speed operation of mechanical logic gates.} (${\bf a}$) and (${\bf b}$) The temporal response of AND and OR gates, schematically depicted in the left panels, in regions G3 and G1 of WG 1 respectively. The two input signals waves are modulated by a PRBS at 3 kb/s, shown in the right panels, along with the idler encoding the logical output. (${\bf c}$) The eye diagram for the AND and OR gates operated in region G2 when the input signals are modulated by a PRBS at 1 kb/s.  The eye openings indicate the successful implementation of mechanical logic gates at high speed in the phonon WG.}
\label{fig 4}
\vspace{-0.5cm}
\end{center}
\end{figure*}

\begin{figure*}[t]
\begin{center}
\vspace{-1.5cm}\hspace{0cm}
\includegraphics[scale=0.9]{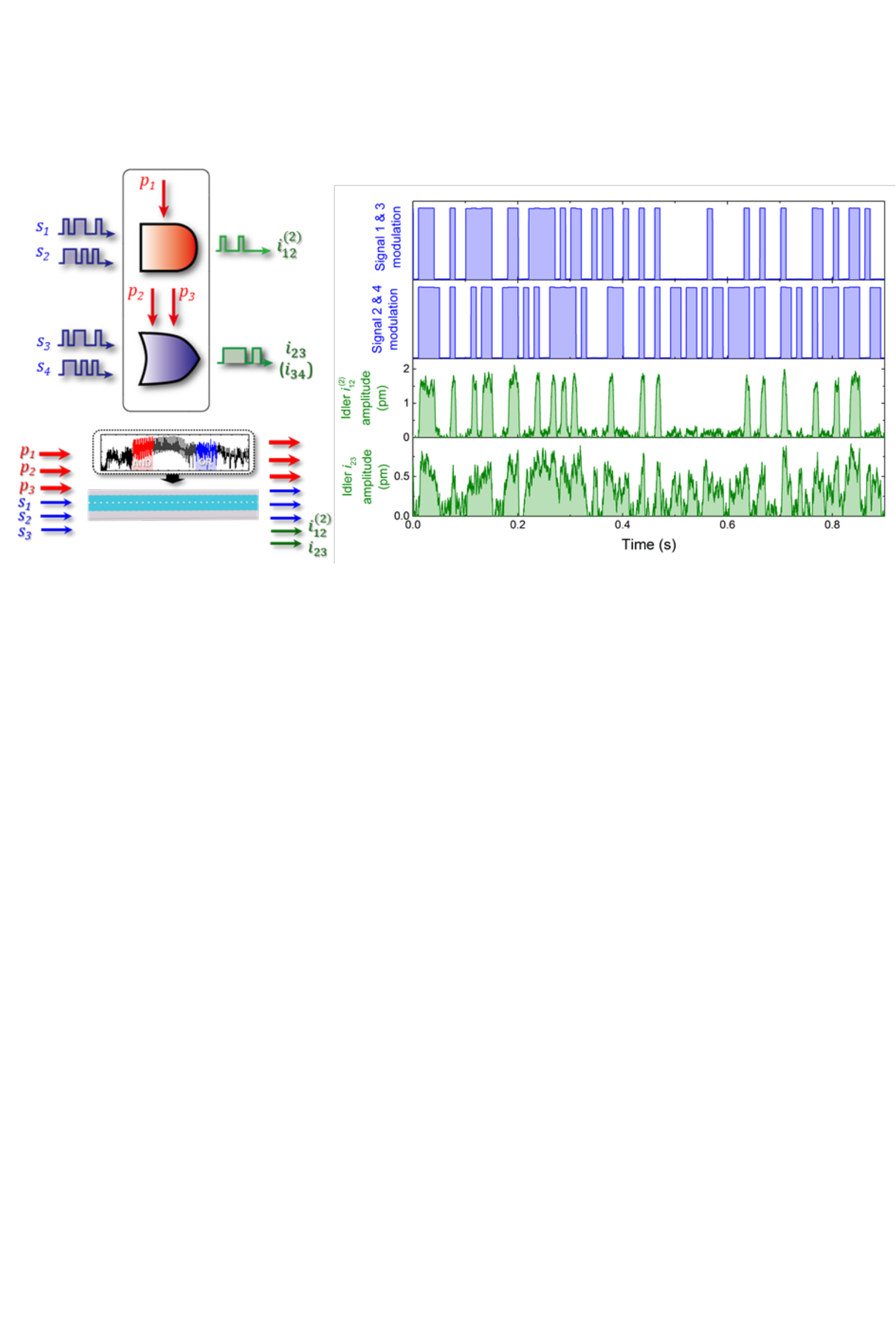}
\vspace{-12cm}
\caption{\textbf{Parallel logic gates in the phonon WG.} The temporal response of both AND and OR gates, depicted schematically in the left panels, when implemented simultaneously in WG 1. The AND gate is executed in G1 with two randomly amplitude modulated input signals ${\it f}_{\rm s1}$ = 2.940 MHz and ${\it f}_{\rm s2}$= 2.990 MHz and one CW pump ${\it f}_{\rm p1}$ = 3.254 MHz yielding the AND idler ${\it f}_{\rm i12}^{\rm (2)}$ = 3.618 MHz. The OR gate is executed in G4 with the two input signals ${\it f}_{\rm s3}$ = 6.020 MHz and ${\it f}_{\rm s4}$ = 6.110 MHz that have the same random amplitude modulation as the AND gate and two CW pumps ${\it f}_{\rm p2}$ =6.315 MHz and ${\it f}_{\rm p3}$ =6.360 MHz which yield an OR idler at ${\it f}_{\rm i23}={\it f}_{\rm i34}$ = 6.610 MHz. The right panels show the temporal response of the randomly amplitude modulated input signals and the resultant output AND, OR logical idlers.}
\label{fig 5}
\vspace{-0.5cm}
\end{center}
\end{figure*}

\setcounter{section}{0} 
\setcounter{figure}{0}
\setcounter{equation}{0}

\renewcommand{\thesection}{Supplementary Note \arabic{section}}
\renewcommand{\thesubsection}{Supplementary Note \arabic{section}.\arabic{subsection}}
\renewcommand{\figurename}{\small{SUPPLEMENTARY FIGURE }\hspace{-0.11cm}}
\renewcommand{\theequation}{S\arabic{equation}}

\onecolumngrid 
\clearpage 

\linespread{1.0}












\section{Broadband frequency conversion via FWM}

\hspace*{1em}In order to demonstrate the availability of broadband frequency conversion via four-wave-mixing (FWM) in the phonon waveguide (WG), the pump wave is fixed on the Fabry-Perot (FP) resonance peak at 4.468 MHz and the signal wave is swept in the transmission band of WG 2 between 3.7-5.2 MHz. The resultant idler phonon wave generated in the transmission band is shown in Supplementary Fig. 1. As the signal frequency passes through the FP resonances, the amplitude of the idler wave is enhanced, due to the field enhancment in the FWM process, and it reproduces the FP resonance spectral structure observed in the transmission measurements.\\

\section{Idler amplification via Fabry-Perot resonance}
\hspace*{1em}The FP resonances in the WG can be used to enhance the efficiency of the mechanical FWM process. Supplementary Fig. 2(a) shows the idler amplitude as a function of pump excitation voltage. The idler (green lines in Supplementary Fig. 2(b)) generation efficiency is strongly dependent on the frequency mismatch of it and the signal and pump waves (blue and red lines in Supplementary Fig. 2(b)) with respect to the FP resonances. For instance when both the signal and pump excitation frequencies are detuned from the FP frequencies (grey points in Supplementary Fig. 2(a) and bottom panel in Supplementary Fig. 2(b)), the idler amplitude is markedly less than when either the signal or pump waves are tuned onto the FP resonances (green and blue points in Supplementary Fig. 2(a) and the middle panels in Supplementary Fig. 2(b) respectively). Finally if both the signal and pump excitations are tuned onto the FP resonances, the idler amplitude exhibits a further amplification of 7 dB as detailed by the red points in Supplementary Fig. 2(a) and the top panel in Supplementary Fig. 2(b). This enhancement in the idler generation is due to the reduced phase mismatch between the signal, pump and idler waves, in addition to energy conservation, which is almost entirely eliminated when all the waves are located at the spectral positions of the FP resonances.\\

\section{Frequency dependence of the primary logic gates}

\hspace*{1em}The broadband response of the idler generation, detailed in Supplementary Note 1, indicates that the primary logic gates executed in the main text are not limited to narrow spectral regions as is the case for mechanical resonator based logic gates. The spectral dependence of the logic gates can be predicted from the details of the FWM process used to generate them and is described below. 
 
\subsection{AND gate}
\hspace*{1em}As detailed in the main text, the AND gate is executed by employing two signal waves into which binary information is encoded and a single pump wave ($\it{f}_{\rm p}$) which generate the 2nd-order idler waves. As a first step to this gate, signal wave 1 ($\it{f}_{\rm s1}$) is activated and signal wave 2 ($\it{f}_{\rm s2}$) is inactive corresponding to the binary input 10. In this configuration an idler from $\it{f}_{\rm s1}$ is generated when the pump is activated as ${\it f}_{\rm i1}^{\rm (1)}$ = 2${\it f}_{\rm p}$ - ${\it f}_{\rm s1}$, namely the idler has a negative variation with respect to the signal wave as depicted in the left panel of Supplementary Fig. 3(a) via the blue, red and green lines (throughout) for the signal, pump and idler waves respectively, and experimentally confirmed in the left panel of Supplementary Fig. 3(b). A similar observation is also made in the opposite configuration, namely the logical input 01 yielding the idler ${\it f}_{\rm i2}^{\rm (1)}$ = 2${\it f}_{\rm p}$ - ${\it f}_{\rm s2}$ as depicted in the middle panel of Supplementary Fig. 3(a) and experimentally confirmed in the middle panel of Supplementary Fig. 3(b). Note that the ${\it f}_{\rm i1}^{\rm (1)}$ and ${\it f}_{\rm i2}^{\rm (1)}$ idlers do not overlap. Finally with both signal waves activated corresponding to the logical input 11, 2nd-order idlers (${\it f}_{\rm i12}^{\rm (2)}$ and ${\it f}_{\rm i21}^{\rm (2)}$) are generated as: 
\begin{eqnarray}
f_{\rm i12}^{\rm (2)} &=& 2f_{\rm i1}^{\rm (1)}-f_{\rm i2}^{\rm (1)} \nonumber \\
&=& 2(2f_{\rm p}-f_{\rm s1})-(2f_{\rm p}-f_{\rm s2})\nonumber \\
&=& 2f_{\rm p}-2f_{\rm s1}+f_{\rm s2},
\end{eqnarray}
and,
\begin{eqnarray}
f_{\rm i21}^{\rm (2)} &=& 2f_{\rm i2}^{\rm (1)}-f_{\rm i1}^{\rm (1)}, \nonumber \\
&=& 2(2f_{\rm p}-f_{\rm s2})-(2f_{\rm p}-f_{\rm s1}) \nonumber \\
&=& 2f_{\rm p}-2f_{\rm s2}+f_{\rm s1}.
\end{eqnarray}
and are depicted by the dark green lines in the right panel of Supplementary Fig. 3(a) and are experimentally confirmed in the right panel of Supplementary Fig. 3(b). Since these 2nd-order idlers only emerge when both signal waves are activated they naturally enable the realisation of the AND gate. Importantly since the 2nd-order idlers wave can be tuned across the entire transmission band in the WG, by adjusting the signal frequency, they enable this gate to be executed over a wide range of frequencies.

\subsection{OR gate}
\hspace*{1em}Again as detailed in the main text, the OR gate is realised by employing two signal waves with ${\it f}_{\rm s2}$ = ${\it f}_{\rm s1}$ + $\Delta$ into which binary information is encoded and two pump waves with ${\it f}_{\rm p2}$ = ${\it f}_{\rm p1}$ + $\Delta$/2 which generate the OR idlers. Specifically activating ${\it f}_{\rm s1}$ corresponding to the logical input 10 yields two idlers with
\begin{eqnarray}
f_{\rm i11} = 2f_{\rm p1}-f_{\rm s1}, \nonumber \\
f_{\rm i21} = 2f_{\rm p2}-f_{\rm s1}.
\end{eqnarray}
which are depicted in the left panel of Supplementary Fig. 4(a) and experimentally confirmed in the left panel of Supplementary Fig. 4(b). Conversely activating ${\it f}_{\rm s2}$ corresponding to the logical input 01 yields the idlers
\begin{eqnarray}
f_{\rm i12} = 2f_{\rm p1}-f_{\rm s2}, \nonumber \\
f_{\rm i22} = 2f_{\rm p2}-f_{\rm s2}.
\end{eqnarray}
which are depicted in the middle of Supplementary Fig. 4(a) and experimentally confirmed in the middle of Supplementary Fig. 4(b). Crucially it can now be shown using equations (S3) and (S4)
\begin{eqnarray}
f_{\rm i22} &=& 2f_{\rm p2}-f_{\rm s2} \nonumber \\
&=& 2(f_{\rm p1}+{\rm \Delta}/2)-(f_{\rm s1}+{\rm \Delta}) \nonumber \\
&=& 2f_{\rm p1}-f_{\rm s1} \nonumber \\
&=& f_{\rm i11}.
\end{eqnarray}
namely these idlers are degenerate. Therefore activating both signal waves corresponding to the logical input 11 yields the idler response depicted in the right panel of Supplementary Fig. 4(a) with the degenerate idlers given by the dark green line which is experimentally confirmed in the right panel of Supplementary Fig. 4(b). Naturally these idlers are always present if either or both signal waves are activated and thus they naturally lead to an OR gate which can be executed over a broad range of frequencies in the phonon WG.\\

\subsection{XOR gate}
\hspace*{1em}As detailed in the main text, the XOR gate is executed in a similar configuration to the OR gate except the pump phase is adjusted so that the degenerate idlers destructively interfere as shown explicitly in Supplementary Fig. 5. Indeed when the phase difference between the two pumps is $\pi$/2 or 3$\pi$/2, the degenerate idlers are eliminated from the destructive interference.\\
\hspace*{1em}By fixing the phase difference between the two pumps at $\pi$/2 and activating ${\it f}_{\rm s1}$ (${\it f}_{\rm s2}$) corresponding to the logical 10 (01) yields the idler response shown in the left (middle) panel of Supplementary Fig. 6(a) which is extracted from equations (S3), (S4) and (S5) with the degenerate idler captured by the dark green line and the pump with the phase difference in the purple line. The corresponding experimental response is shown in the left (middle) panel of Supplementary Fig. 6(b). Finally activating both signal waves corresponding to the logical input 11 yields the idler response depicted in the right panel of Supplementary Fig. 6(a) with the degenerate idler being eliminated which is experimentally confirmed in the right panel of Supplementary Fig. 6(b). Naturally this idler encapsulates the XOR gate as it is absent when both signal waves are activate and it can be executed over a broad range of frequencies in the phonon WG.

\newpage

\begin{figure*}[h]
\begin{center}
\vspace{-0cm}\hspace{0cm}
\includegraphics[scale=0.6]{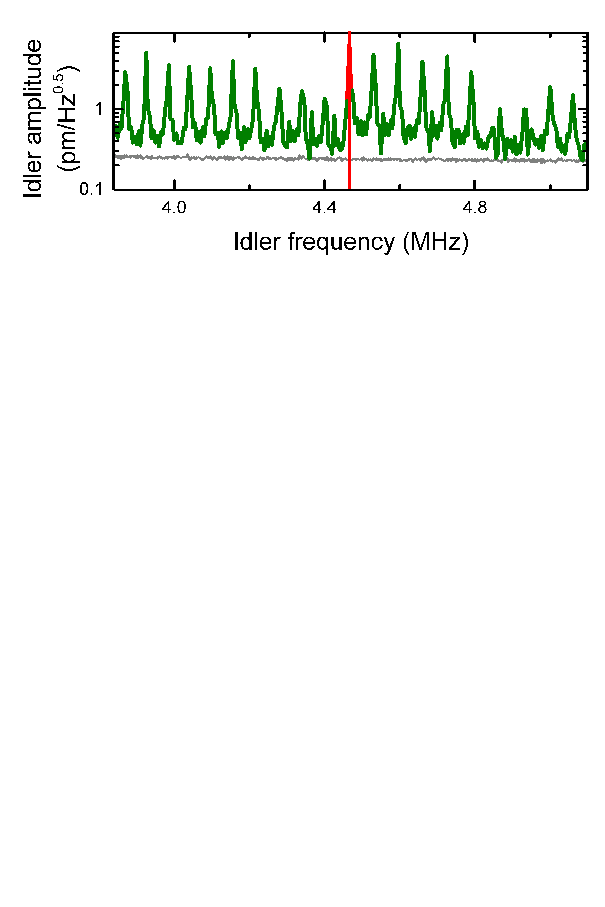}
\vspace{-10.0cm}
\caption{\textbf{Broadband idler generation.}  The spectral dependence of the idler waves (green) generated by sweeping the signal wave from 3.7 to 5.2 MHz with an amplitude of 0.35 V$_{\rm rms}$ whilst the pump wave is excited at a fixed frequency of  4.468 MHz (red line) and an amplitude of 1.0 V$_{\rm rms}$ in WG 1. The grey line shows the background noise in the WG measured via optical interferometry and demodulated in a spectrum analyser in the absence of FWM.}
\label{fig S3}
\vspace{0.5cm}
\end{center}
\end{figure*}

\begin{figure*}[h]
\begin{center}
\vspace{-0cm}\hspace{0cm}
\includegraphics[scale=0.6]{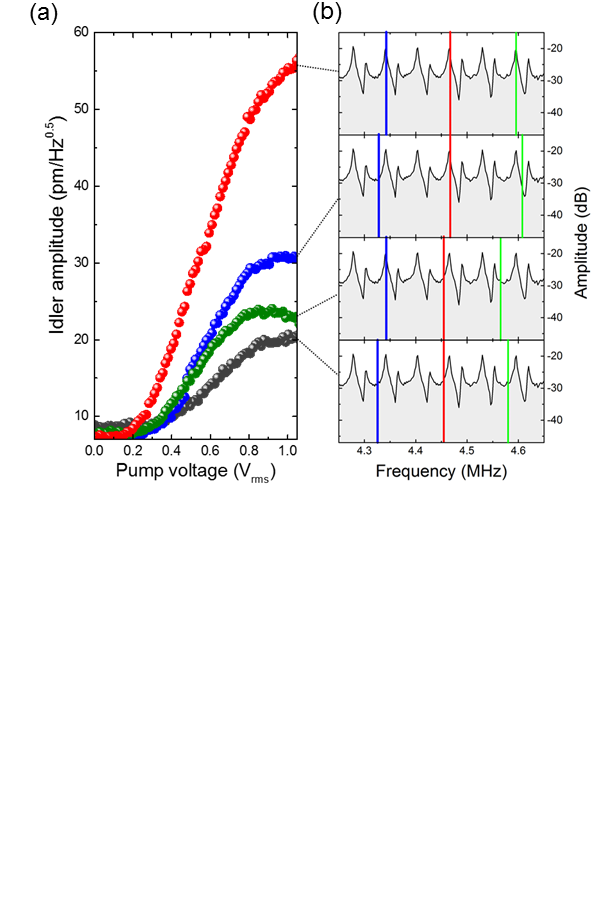}
\vspace{-6cm}
\caption{\textbf{The pump voltage dependence of the idler amplitude.} ($\bf{a}$) The pump amplitude dependence of the idler amplitude for various combinations of signal and pump excitation frequencies that are detailed in ($\bf{b}$). ($\bf{b}$) The peak structure corresponds to the FP resonances in the WG, acquired from the transmission measurement in the upper panel of Fig. 2, and all measurements were conducted with a signal amplitude of 0.35 V$_{\rm rms}$.}
\label{fig S1}
\vspace{-0.5cm}
\end{center}
\end{figure*}

\begin{figure*}[h]
\begin{center}
\vspace{-0cm}\hspace{8cm}
\includegraphics[scale=0.7]{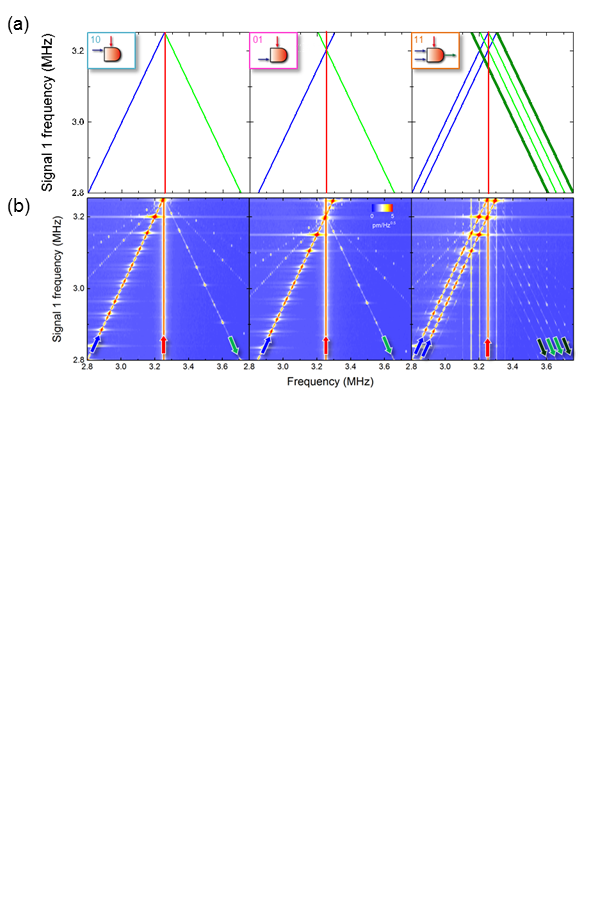}
\vspace{-9cm}
\caption{\textbf{AND gate.} ($\bf{a}$) The spectral dependence of the idler waves (green and dark green lines), as detailed above, generated from a single pump wave (red line)  at fixed frequency as function of two signal waves (blue lines) when either is on/off (left panel) or off/on (middle panel) or when both are on (right panel) corresponding to logical inputs 10, 01 and 11 respectively. In the latter configuration, 2nd-order idlers (dark green lines) are also generated as detailed in equations (S1) and (S2) which are used to execute the AND gate. ($\bf{b}$) The experiments were conducted in region G1 of WG 1 with the 1.5 V$_{\rm rms}$ pump wave fixed at 3.254 MHz (red arrows). The signals waves (blue arrows) are injected with an amplitude of 1.5 V$_{\rm rms}$ with ${\it f}_{\rm s2}$ = ${\it f}_{\rm s1}$ + 50 kHz in the frequency range of 2.80-3.25 MHz. The 2nd-order idler waves (dark green arrows) are generated between 3.25-3.75 MHz only in the presence of both signals (right panel), enabling the AND gate to be realised in the phonon WG. The insets show the signal and pump configurations of the logic gate being performed.}
\label{fig S3-1}
\vspace{0.5cm}
\end{center}
\end{figure*}

\begin{figure*}[h]
\begin{center}
\vspace{-0cm}\hspace{8cm}
\includegraphics[scale=0.7]{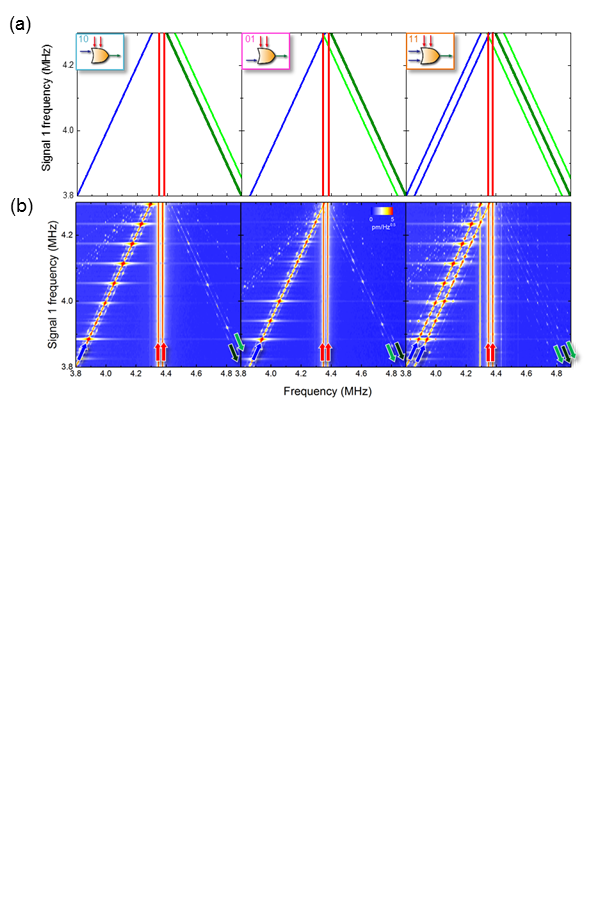}
\vspace{-9cm}
\caption{\textbf{OR gate. } ($\bf{a}$) The spectral dependence of the idler waves (green and dark green lines), as detailed above, generated from two pump waves (red lines)  at fixed frequencies as function of two signal waves (blue lines) when either is on/off (left panel) or off/on (middle panel) or when both are on (right panel) corresponding to logical inputs 10, 01 and 11 respectively. In the latter configuration, degenerate idlers (dark green line) are present as detailed in equations (S5) which are used to execute the OR gate. ($\bf{b}$) The experiments were conducted in region G2 of WG 1 with two 1 V$_{\rm rms}$ pump waves fixed at ${\it f}_{\rm p1}$ = 4.347 MHz and ${\it f}_{\rm p2}$ = 4.375 MHz i.e. $\Delta$/2 = 28 kHz (red arrows). Two signals waves (blue arrows) are injected with an amplitude of 1.5 V$_{\rm rms}$ at ${\it f}_{\rm s2}$ = ${\it f}_{\rm s1}$ + 56 kHz and in the frequency range of 3.8-4.3 MHz (blue arrows). The degenerate idler waves ${\it f}_{\rm i11}$ = ${\it f}_{\rm i22}$ (dark green line) are always observed between 4.5-4.9 MHz when either or both signal waves are activate, yielding the OR gate in the phonon WG. The insets show the signal and pump configurations for the logic gate being performed.}
\label{fig S4}
\vspace{0.5cm}
\end{center}
\end{figure*}

\begin{figure*}[h]
\begin{center}
\vspace{-0cm}\hspace{0cm}
\includegraphics[scale=0.6]{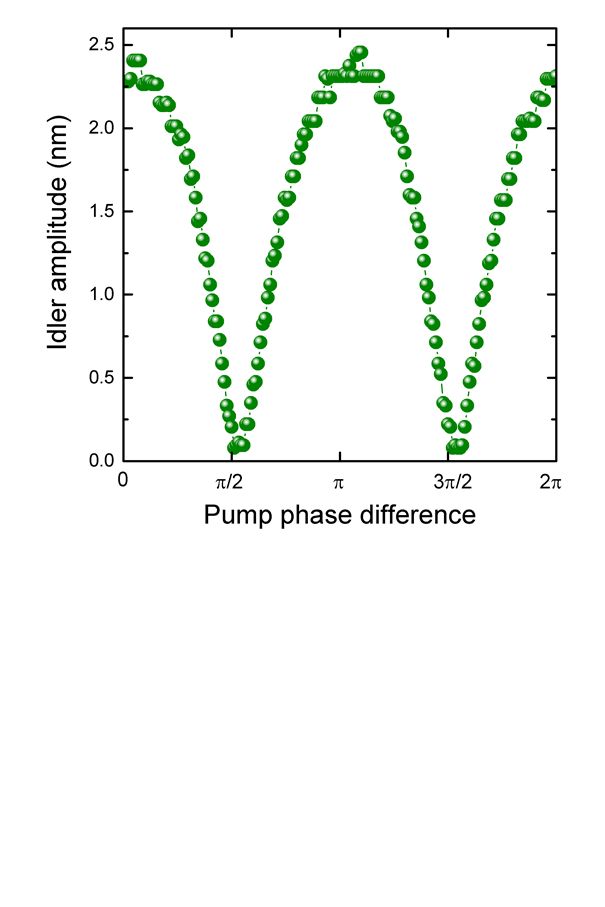}
\vspace{-5.5cm}
\caption{\textbf{The pump phase dependence of the spectrally degenerate idler wave.} The degenerate idler amplitude at ${\it f}_{\rm i11}={\it f}_{\rm i22}$= 5.687 MHz when the phase difference between ${\it f}_{\rm p1}$ = 5.356 MHz and ${\it f}_{\rm p2}$ = 5.390 MHz with amplitudes of 1.5 V$_{\rm rms}$ whilst both signal waves are active at ${\it f}_{\rm s1}$ = 5.025 MHz and ${\it f}_{\rm s2}$ = 5.093 MHz with amplitudes of 2.0 V$_{\rm rms}$. The degenerate idlers undergo interference when the pump phase is adjusted and the destructive interference at $\pi$/2 or 3$\pi$/2 can be exploited to build an XOR gate when both signal waves are active.}
\label{fig S3-4}
\vspace{0.5cm}
\end{center}
\end{figure*}

\begin{figure*}[h]
\begin{center}
\vspace{-0cm}\hspace{8cm}
\includegraphics[scale=0.7]{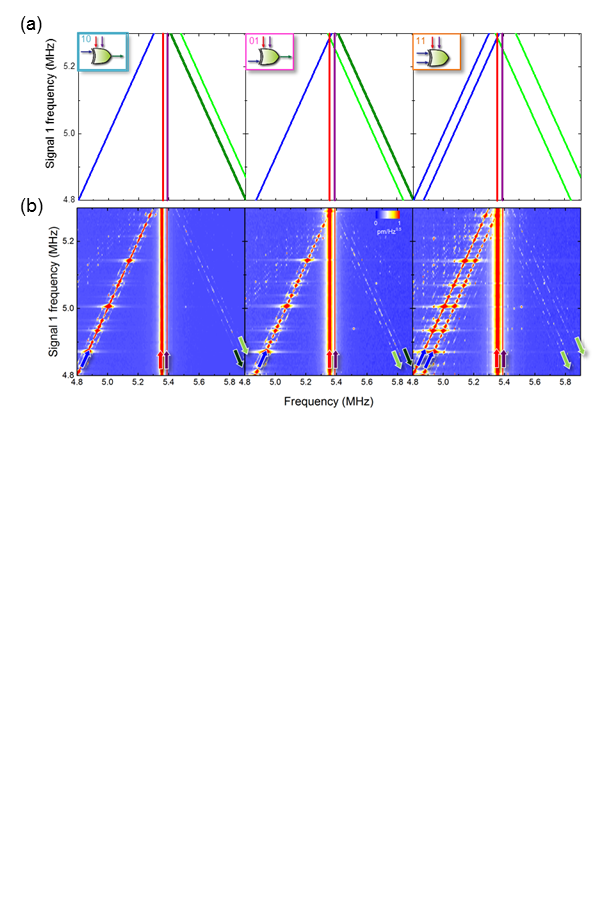}
\vspace{-9cm}
\caption{\textbf{XOR gate.} ($\bf{a}$) The spectral dependence of the idler waves (green and dark green lines), as detailed above, generated from two pump waves with $\pi$/2 phase difference (red and purple lines respectively) at fixed frequencies as function of two signal waves (blue lines) when either is on/off (left panel) or off/on (middle panel) or when both are on (right panel) corresponding to logical inputs 10, 01 and 11 respectively. In the latter configuration, the degenerate idler (dark green line) undergoes destructive interference as detailed in Supplementary Fig. 5 and is eliminated thus yielding an XOR gate. ($\bf{b}$) The experiments were conducted in region G3 of WG 1 with two 1.5 V$_{\rm rms}$ pump waves fixed at ${\it f}_{\rm p1}$ = 5.356 MHz and ${\it f}_{\rm p2}$ = 5.390 MHz i.e. $\Delta$ = 67.5 kHz with a $\pi$/2 phase shift (red and purple arrows). Two signals waves (blue arrows) are injected with an amplitude of 2.0 V$_{\rm rms}$ at ${\it f}_{\rm s2}$ = ${\it f}_{\rm s1}$ + 67.5 kHz in the frequency range of 4.8-5.3 MHz encoding to the logical inputs. The degenerate idler waves at ${\it f}_{\rm i11}$ = ${\it f}_{\rm i22}$ (dark green arrow) observed between 5.45-5.9 MHz is eliminated when both signal waves are activated yielding the XOR gate in the phonon WG. The insets show the signal and pump configurations for the logic gate being performed.}
\label{fig S5}
\vspace{0.5cm}
\end{center}
\end{figure*}






\end{document}